\documentclass[aps,prb,twocolumn,showpacs,superscriptaddress,groupedaddress]{revtex4-1} 
\usepackage{dcolumn}   
\usepackage{bm}        
\usepackage{amssymb}   
\usepackage{amsmath}
\usepackage{natbib}
\usepackage{float}
\usepackage[dvips,final]{graphicx}
\usepackage[tight,TABTOPCAP]{subfigure}
\usepackage{pslatex}
\usepackage{relsize}
\usepackage{bm}
\usepackage{xspace} 

\usepackage{xcolor}
\usepackage{soul}

\def\ba{\begin{eqnarray}}
\def\ea{\end{eqnarray}}
\def\be{\begin{equation}}
\def\ee{\end{equation}}

\bibpunct{[}{]}{,}{n}{}{} 

\begin{document}
\title{Skyrmion bound state and dynamics in an antiferromagnetic bilayer racetrack}

\author{R. C. Silva}
\affiliation{Departamento de Ci\^{e}ncias Naturais, Universidade Federal do Esp\'{i}rito Santo, Rodovia Governador M\'{a}rio Covas, Km 60, 29932-540, S\~{a}o Mateus, ES, Brazil.}
\author{R. L. Silva}
\affiliation{Departamento de Ci\^{e}ncias Naturais, Universidade Federal do Esp\'{i}rito Santo, Rodovia Governador M\'{a}rio Covas, Km 60, 29932-540, S\~{a}o Mateus, ES, Brazil.}
\author{V. L. Carvalho-Santos}
\affiliation{Departamento de F\'{i}sica, Universidade Federal de Vi\c{c}osa, Avenida Peter Henry Rolfs s/n, 36570-900, Vi\c{c}osa, MG, Brazil}
\author{W. A. Moura-Melo}
\affiliation{Departamento de F\'{i}sica, Universidade Federal de Vi\c{c}osa, Avenida Peter Henry Rolfs s/n, 36570-900, Vi\c{c}osa, MG, Brazil}
\author{A. R. Pereira}
\affiliation{Departamento de F\'{i}sica, Universidade Federal de Vi\c{c}osa, Avenida Peter Henry Rolfs s/n, 36570-900, Vi\c{c}osa, MG, Brazil}


\begin{abstract} 
We investigate the dynamics of two skyrmions lying in distinct layers of an antiferromagnetic bilayer system, consisting of nanostripes with the shape of racetracks. The top and bottom nanostripes are separated by a height offset and they are coupled through a ferromagnetic exchange, allowing the interaction between the skyrmions from both layers. Depending on the distance between the skyrmions they attract each other sufficiently to achieve a bound-state. We also analyze their dynamics when an electric current is applied in a unique layer and we determine how the bound-state nucleation depends on the current density and vertical distance between the skyrmions. Finally, we analyzed the robustness of the bound-states by considering two situations: 1) a system constituted by clean or homogeneous antiferromagnetic racetracks; 2) a system in which randomly distributed magnetic impurities in both layers are included in the system.
\end{abstract}

\flushbottom \maketitle

\thispagestyle{empty}

\section{Introduction}

In the 1960's, Skyrme proposed that the stability of hadrons could be understood from the existence of quantized topological defects in high energy physics \cite{Skyrme1, Skyrme2}. These topological objects are currently known as skyrmions, and this idea was expanded to condensed matter systems, becoming a highly relevant topic in magnetism. The theme of topological structures was elegantly introduced in magnetic materials by Belavin and Polyakov (BP) \cite{BP} as possible solutions for the two-dimensional non-linear sigma model (the continuum limit of the isotropic Heisenberg model). A magnetic skyrmion is a non-collinear spin texture with a whirling magnetic structure where the spins point in all directions, wrapping the internal space of the spin sphere. How many times such a wrapping takes place is a topological invariant and endows skyrmion its topological charge (an integer number). Owing this charge, a skyrmion has topological stability, in such a way that its charge imposes an energy barrier protecting them from decaying to the ground state. In the BP's model, they appear as excited states and their nucleation is not easy to be achieved in magnetic materials described uniquely by the exchange interaction. On the other hand, in the last decade, the possibility of nucleating skyrmions from the competition between the exchange, anisotropy, and Dzyaloshinskii-Moriya (DMI) interactions \cite{Dzyaloshinskii, Moriya, ANBogdonov, ABogdanov, Robler} yielded a large interest in studying such particle-like magnetic objects. 

The increasing interest in magnetic skyrmions is also explained because they are promising candidates for future spintronic applications, especially for low-energy consumption and stable high-density information storage. Moreover, they present important properties for producing well-designed devices in the area of nanotechnology. Amongst these properties, we can highlight their small sizes \cite{Heinze, Wiesendanger, Soumyanarayanan}, the possibility of creating, manipulating, and annihilating them by purely electric means \cite{Iwasaki, Romming, Chouk, Heil, XZhang3}, their topological stability that allows these structures to be observed even at room temperature \cite{Brandao, Woo1}, and the prospect of guiding them under the action of a low-density spin-polarized current \cite{Yu}. Nevertheless, technological applications demanding the skyrmion motion confront the undesired obstacle brought about by the Magnus Force, which deviates the skyrmion trajectory along the direction normal to the external current. This phenomenon is called skyrmion Hall effect (SHE)  \cite{JZang, Jiang, Litzius}. A number of proposals has appeared to bypass such an effect, for instance, the use of engineered materials \cite{Toscano, Silva}, the substitution of skyrmions by other topological magnetic structures as information carriers \cite{Kolesnikov}, and the coupling of skyrmions in magnetic bilayers \cite{XZhang1, Koshibae, Hrabec, Vagson, Cacilhas}. 

Another alternative way of eliminating SHE is by considering antiferromagnetic (AFM) systems. Recent advances in electrical manipulation and detection of AFM systems have created a new branch in spintronics: the topological antiferromagnetic spintronics \cite{Smejkal, XZang-JAP}. In AFM systems, the whole lattice can be viewed as two ferromagnetic (FM) interspersed sublattices. Therefore, the sublattices topological charges are opposite each other, yielding a Magnus force that acts in opposite directions on each sublattice,  canceling the SHE on a AFM skyrmion. Consequently, skyrmions can move in straight lines along with the applied electric current without deflection \cite{Barker, XZhang2}. There are other advantages for considering skyrmions in AFM media, such as (i) their velocity is dozens of times greater than the FM skyrmions under the same electric current density; (ii) the minimum current density needed to put the AFM skyrmion in motion is about a hundred times smaller than that used to put the FM counterparts in movement \cite{Jin}; (iii) external magnetic fields applied to the sample do not have any effect, making the AFM system more rigid against magnetic perturbations \cite{Barker}. Recently, skyrmions have been observed in GdFeCo and DyCo$_{3}$ ferrimagnetic films \cite{Woo2, Chen}, and a fractional AFM skyrmion lattice was experimentally reported in MnSc$_{2}$S$_{4}$ \cite{Gao}. Indirect evidence of the nucleation of AFM skyrmions was also disclosed in Ref. \cite{Raicevic}.
\begin{figure}[htb]
    \begin{center}
	\subfigure[]{\includegraphics[width=8.10cm]{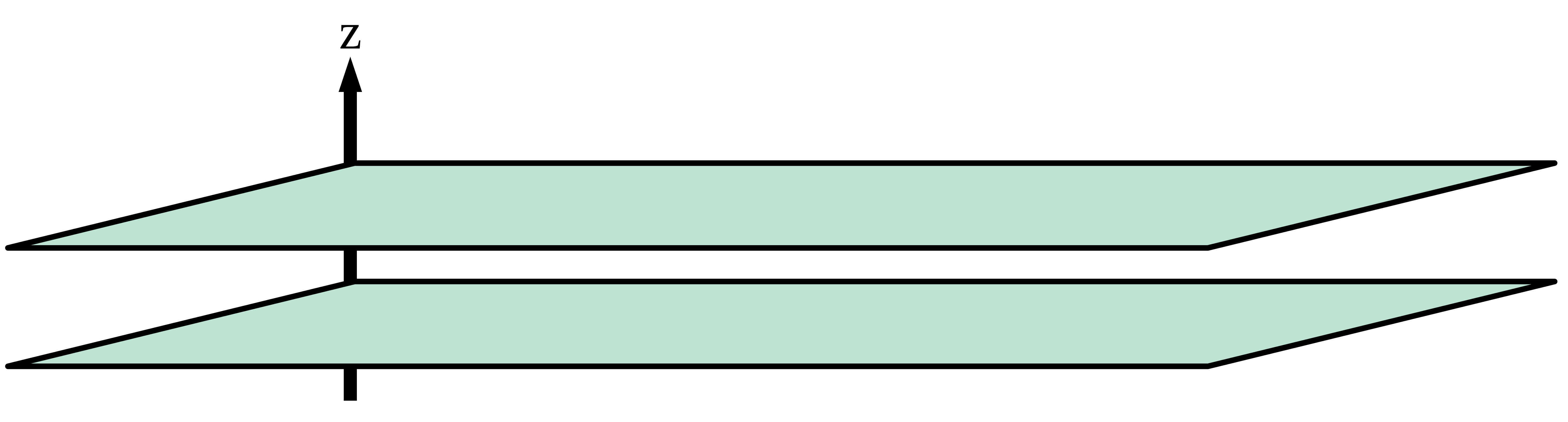}}
	\vskip 0.2cm
	\subfigure[]{\includegraphics[width=8.0cm]{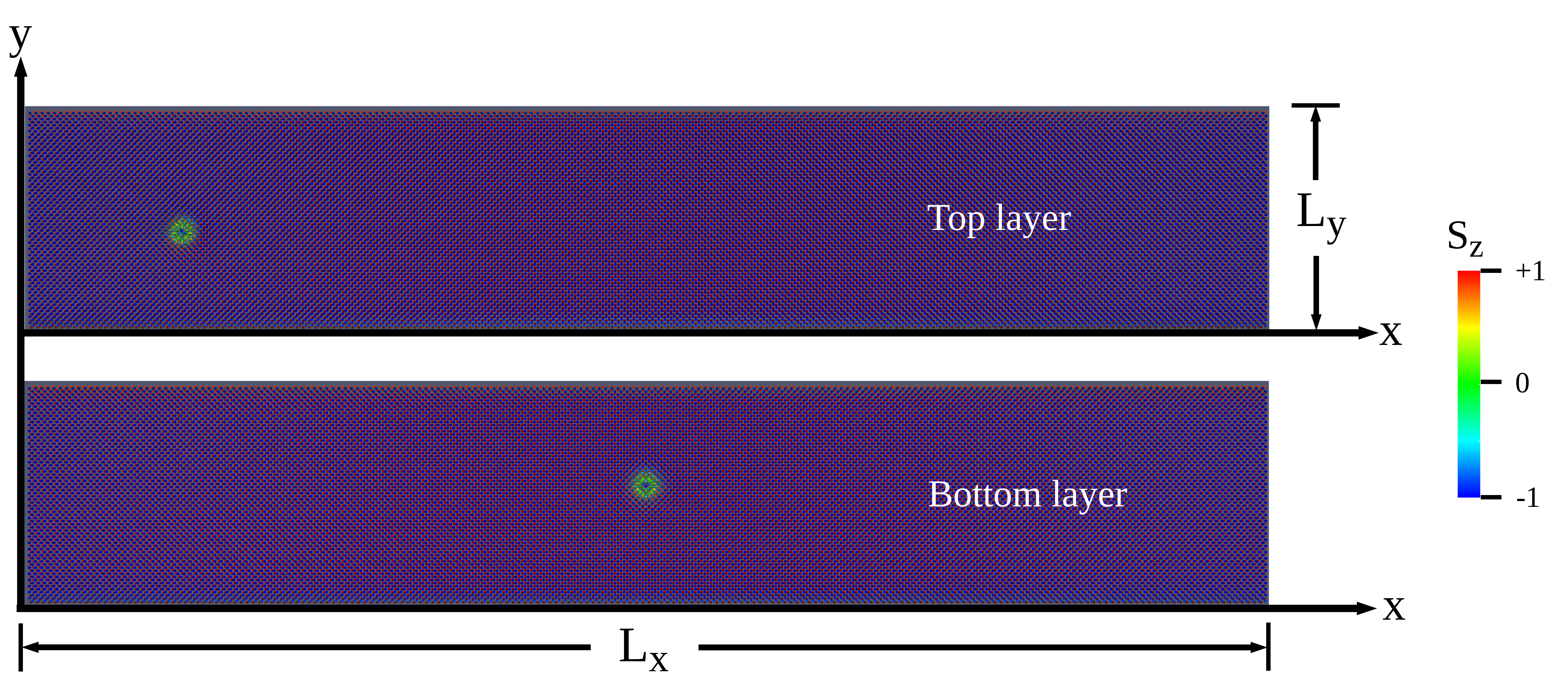}}
	\caption{(Color online) (a) Lateral view of the bilayer system consisting by two parallel racetracks separated by a height offset along $z$ direction. (b) Side-by-side view of the top and bottom racetracks with their respective skyrmions in the initial positions.}
	\label{bilayer1}
    \end{center}
\end{figure}

In this work, we are interested in studying the dynamical properties of interfacial AFM skyrmions in a bilayer system, as schematically shown in the Fig. \ref{bilayer1}. The bilayer consists of two parallel AFM racetracks separated by a height offset which is filled by an insulating spacer. In addition, each layer comprises a unique skyrmion. In this context, we consider that exchange-like interactions can mediate the coupling between the layers \cite{Burgler, Orozco}, where its strength decays exponentially with separation between the layers (or the thickness of the spacer) \cite{Bruno}. Several experimental works \cite{Romming, SHYang, GChen, Dupe,Luchaire, Matsuno, Parkin, Parkin2, Yuasa, Nunn} showed the possibility to manipulate the interlayer interaction artificially, where the ferro- or antiferromagnetic coupling is selected by inserting a magnetic metallic spacer with a controlled chemical composition. Our results evidence that depending on the distance between the skyrmions a bound-state of them is achieved. When an external electric current is applied in the top layer, its skyrmion drags that lying in the bottom layer and they displace as a single object. Simulation for a large range of parameters have been performed yielding a state diagram showing that the nucleation of skyrmion bound-states depends on the current density and vertical distance between the skyrmions. Finally, we analyzed their dynamics in the presence of magnetic impurities in the racetracks. It is noticed that due to the interaction between the skyrmion bound-state and magnetic impurities, they experience a slight deviation from their rectilinear trajectories.

\section{Model and Methods}

In order to describe the AFM bilayer racetrack, we consider two planar and identical layers parallel to each other and separated by a height offset. Each layer is a two-dimensional square lattice whose magnetization is represented by an array of unit dimensionless vector $\vec{S}_{i} = \left(S^{x}_{i},S^{y}_{i},S^{z}_{i}\right)$, accounting for the magnetic moment located at the site \textit{i}. In our simulations, we have used nanostripes with length L$_{x}$ = 400 \textit{a} and width  L$_{y}$ = 70 \textit{a}, where \textit{a} is the lattice constant of the layers. Thus, our system comprises $2$L$_{x}$L$_{y}$ magnetic moments, interacting by means of the Hamiltonian below:
\begin{equation} \label{hamiltonian}
	\begin{array}{ccc}
		\displaystyle{\mathcal{H}} &=& \displaystyle{\sum_{m=B,T}\left[\sum_{\langle i,j \rangle}J_{ij,m} \vec{S}_{i,m} \cdot \vec{S}_{j,m} - \sum_{\langle i,j \rangle}\vec{D}_{ij,m} \cdot \left(\vec{S}_{i,m} \times \vec{S}_{j,m}\right) \right.}\\
		\\
		& & \displaystyle{\left.-\sum_{i}K_{i,m}\left(\vec{S}_{i,m} \cdot  \hat{e}_{k}\right)^{2} \right]}-\displaystyle{\sum_ {i}J_{i,BT}\, \vec{S}_{i,B} \cdot \vec{S}_{i,T}},
	\end{array}	
\end{equation}
\noindent 
where $J_{ij,m}=J=+1$ is the intralayer antiferromagnetic constant, $\vec{D}_{ij,m}$ is the Dzyaloshinskii-Moriya (DM) vector, $K_{i,m}>0$ is the easy-axis anisotropy constant since we take $\hat{e}_{k}=\hat{z}$, and  $J_{i,BT}$ is the interlayer ferromagnetic constant \cite{XZhang1}. The sum over $\langle i,j \rangle$ is taken over all pairs of nearest neighbor magnetic moments lying in the same layer, and $m=(B,T)$ indicates the layer index ($B$ for the bottom layer and $T$ for the top one). In this context, first, second, and third terms in Eq. (\ref{hamiltonian}) describe the intralayer interactions given by the exchange, DM, and anisotropy terms, respectively. Here, we consider $\vec{D}_{ij,m}=D_{ij,m}\left(\hat{r}_{ij,m}\times \hat{z}_{m}\right)$, where $\hat{z}_{m}$ is a normal unitary vector of each racetrack, and $\hat{r}_{ij,m}$ is the unitary vector pointing from site \textit{i} to site \textit{j} \cite{Rohart, Yang}. In our simulations, we set $D_{ij,m}=0.30 J$ and $K_{i,m}=0.04 J$, which favor the appearance of N{\'e}el skyrmions (hedgehog-type texture). The fourth term in Eq. (\ref{hamiltonian}) describes the interlayer exchange interaction. Here, we consider $J_{i,BT} = 0.015 J>0$, resulting in a ferromagnetic coupling between the magnetic moments in layers T and B. A detailed study about the skyrmion stability in an AFM bilayer as a function of magnetic parameters was performed in Ref. \cite{Ubiergo}. We have used mixed boundary conditions, that is, periodic boundary condition along $x$, whereas on \textit{y}-axis open boundary condition is taken.

The magnetization dynamics is determined by solving the Landau-Lifshitz-Gilbert (LLG) equation \cite{Landau, Gilbert}, augmented by the Zhang-Li spin-transfer torque term \cite{Zhang-Li}, as below:
\begin{equation} \label{LLGZL}
	\begin{array}{ccc}
		\displaystyle{\frac{d \vec{S}_{i,m}}{dt}} = \displaystyle{-\gamma \vec{S}_{i,m}\times \hat{b}_{i,m} -\alpha \vec{S}_{i,m} \times \frac{d\vec{S}_{i,m}}{dt}}\\
		\\
		\displaystyle{-\frac{\nu \delta_{mT}}{M_{sat}} \vec{S}_{i,T} \times \left[ \vec{S}_{i,T}\times \left(\hat{j}_{e} \cdot \nabla \right)\vec{S}_{i,T}\right] -\frac{\beta \nu \delta_{mT}}{M_{sat}}\vec{S}_{i,T} \left(\hat{j}_{e} \cdot \nabla\right) \vec{S}_{i,T}},
	\end{array}	
\end{equation}
\noindent
where $\gamma$ is the gyromagnetic ratio, $\hat{b}_{i,m} = -\frac{1}{\mu_{S}}\frac{\partial \mathcal{H}}{\partial \vec{S}_{i,m}}$ is the local effective field at the lattice site \textit{i}, and $\alpha$ is the Gilbert damping parameter. The last two terms in Eq. (\ref{LLGZL}) refer to the electric spin-polarized current, which is considered being applied only in the top layer, represented by the Kronecker delta, $\delta_{mT}$. Additionally, $\nu = \frac{P j_{e} \mu_{B}}{e M_{sat} \left(1+\beta^{2}\right)}$, $P$ is the degree of polarization of the spin current, $\mu_{B}$ is the Bohr magneton, $e$ is the modulus of the electronic charge, $M_{sat}$ is the saturation magnetization, $\beta$ is the degree of non adiabadicity, and $j_{e}$ is the norm of the electric current density vector. Here, we consider an electric current density $\vec{j}_{e} = -j_{e}\,\, \hat{x}$ in order to move the AFM skyrmion of the top layer from the left to the right. In our simulations we set  $\alpha = 0.15$, $\beta = 0.08$,  $\gamma =1.0$, and $P = 0.90$. The LLG equation has been integrated using a fourth-order Runge-Kutta method with a time step of $\Delta \tau = 0.0001$ (measured in units of $\hbar/J\sim 10^{-13} {\rm s}$, for typical samples).

As an initial condition we take an AFM skyrmion given by the Belavin-Polyakov configuration, as follows: 
\begin{equation} \label{BPskyrmion}
	\begin{array}{l}
		\displaystyle{S_{xy}^{x}} = \displaystyle{\left(-1\right)^{x+y}\frac{\left(x-x_{sk,m}\right)R}{\left( x-x_{sk,m}\right)^{2}+\left( y-y_{sk,m}\right)^{2}+R^{2}}} \\
		\\ 
		\displaystyle{S_{xy}^{y}} = \displaystyle{\left(-1\right)^{x+y}\frac{\left(y-y_{sk,m}\right)R}{\left( x-x_{sk,m}\right)^{2}+\left( y-y_{sk,m}\right)^{2}+R^{2}}} \\
		\\ 
		\displaystyle{S_{xy}^{z}} = \displaystyle{\left(-1\right)^{x+y}\frac{\left( x-x_{sk,m}\right)^{2}+\left( y-y_{sk,m}\right)^{2}-R^{2}}{2\left[\left( x-x_{sk,m}\right)^{2}+\left( y-y_{sk,m}\right)^{2}+R^{2}\right]}}\,, \\
		\\ 
	\end{array}	
\end{equation}
\noindent
which represents a single skyrmion with radius \textit{R}, whose center of mass is placed at $\left(x_{sk,m},y_{sk,m} \right)$. To ensure a non-interacting skyrmions state in the initial configuration of the bilayer, the textures lying in different racetracks are initially placed quite far each other. This initial configuration is then relaxed by solving Eq. {\ref{LLGZL}} without the spin-polarized current term. Therefore, the magnetization evolves to the minimum energy configuration, enabling the adjustment of the skyrmion radius and its in-plane magnetization configuration.
 
The position of the skyrmion mass center is calculated using the density of topological charge, defined by \cite{Moutafis}:
 \begin{equation}
	\rho_{k,m} = \dfrac{1}{2} \varepsilon_{ij} \vec{\eta}_{k,m} \cdot \left(\partial_{j} \vec{\eta}_{k,m} \times \partial_{i} \vec{\eta}_{k,m}\right),
 \end{equation}
\noindent
where \textit{i} and \textit{j} sum over the horizontal and vertical directions of $m$-layer, and $\displaystyle{\vec{\eta}_{k,m} = \frac{\vec{S}_{2k,m}-\vec{S}_{2k+1,m}}{2}}$ is the N{\'{e}}el vector. Thereby, the position of the skyrmion center of mass in the $m$-layer, $\vec{r}_{sk,m} = (x_{sk,m},y_{sk,m})$, is given by \cite {Shen, Stier}: 
 \begin{equation}
	\displaystyle{\vec{r}_{sk,m} = \frac{\int \vec{r}_{m} \rho_{m} d^{2}r}{N_{sk,m}}}
 \end{equation}
\noindent
where $\vec{r}_{m} = x_{m}\,\, \hat{x} + y_{m} \,\, \hat{y}$ is the lattice site coordinates of the $m$-layer and $N_{sk,m} = \int \rho_{m} d^{2}r$ is the N{\'{e}}el topological number. The skyrmion radius $\lambda_{m}$ is defined by:
 \begin{equation}
	\displaystyle{\lambda_{m} = \sqrt{\int \left|\vec{r}_{m}-\vec{r}_{sk,m}\right|^{2}\frac{\rho_{m}}{N_{sk,m}}}}
 \end{equation}

\section{Results and Discussion}

\subsection{Bilayer without impurities}

\begin{figure}[htb]
	\begin{center}
		\includegraphics[width=6.50cm]{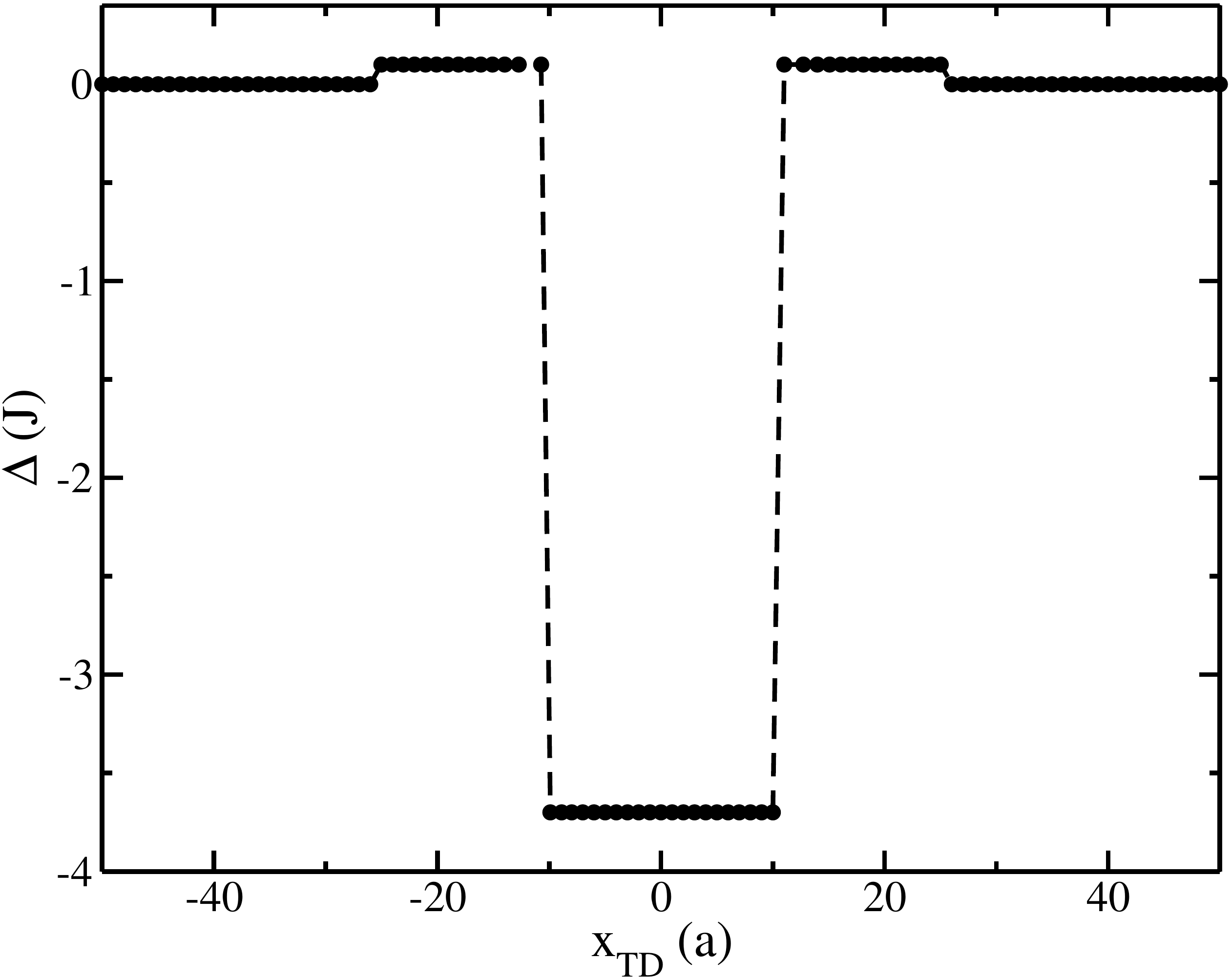}
		\caption{Interaction potential $\Delta$ between skyrmions lying in top and bottom layers as a function of their horizontal distance $x_{TD}$. Namely, note that it is strongly attractive for relatively short separation.}
		\label{potential}
	\end{center}
\end{figure}

Firstly, we have considered that the AFM racetracks composing the bilayer system do not have inhomogeneities or magnetic impurities. In this case, we analyze the possibility of creating a bound-state of skyrmions due to the ferromagnetic coupling between the two layers. By fixing the bottom skyrmion at the middle point of its nanotrack and placing the top skyrmion in different horizontal positions, we have determined the system energy as a function of the horizontal distance between the skyrmions. It is obtained by relaxing the initial magnetization profile, integrating the LLG equation without electrical currents. In this way, we have defined $\Delta = E(x_{TD})-E(\infty)$ as the potential interaction between the top and bottom skyrmions, where $E(x_{TD})$ is the system energy when the horizontal distance between the textures is $x_{TD}$, and $E(\infty)$ consists of the free-skyrmions system energy, when the distance between the skyrmions is very large. Figure \ref{potential} depicts the behavior of $\Delta$ as a function of $x_{TD}$. One can notice that the skyrmions interaction is appreciable only for relatively short distances. Indeed, $\Delta\approx0$ for $|x_{TD}| \geqslant 26 a$, evidencing that the system energy is practically constant and the non-interacting skyrmion state is observed. Nevertheless, if $11 a \leq |x_{TD}| \leq 25 a$, a slightly repulsion between skyrmions takes place whereas for $|x_{TD}| \leq 10 a$ they start to attract each other. Therefore, for sufficiently short separation skyrmions strongly couple each other yielding a bound-state mediated by the intralayer exchange interaction.

\begin{figure}
    \centering
    \includegraphics[width=6.5cm]{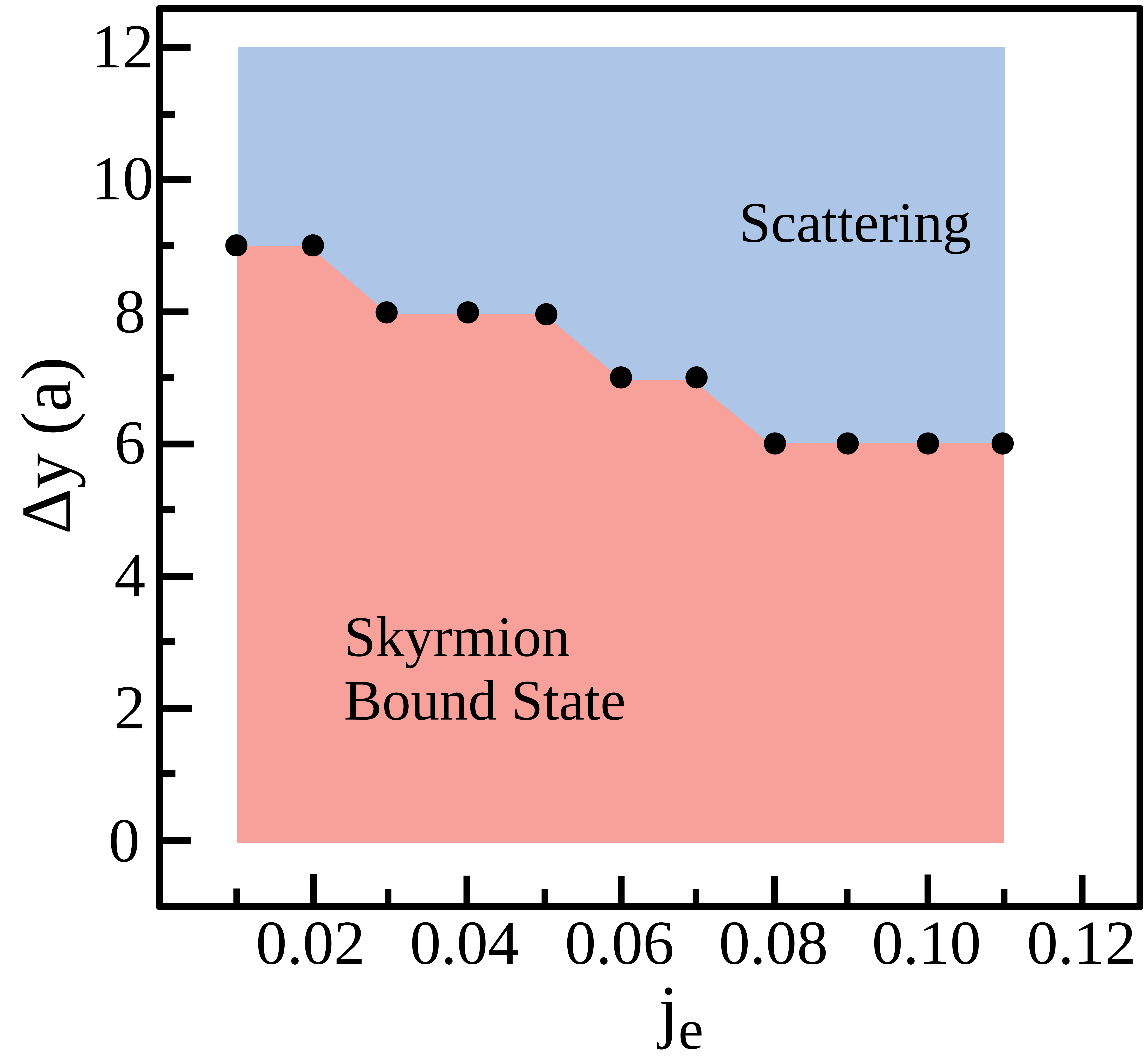}
    \caption{(Color online) Skyrmions may nucleate into bound state or are scattered depending on vertical separation ($\Delta\,y$) and applied  current ($j_e$).}
    \label{fig3New}
\end{figure}

\begin{figure}[htb] \label{Fig4}
   \begin{center}
	\subfigure[]{\includegraphics[width=8.0cm]{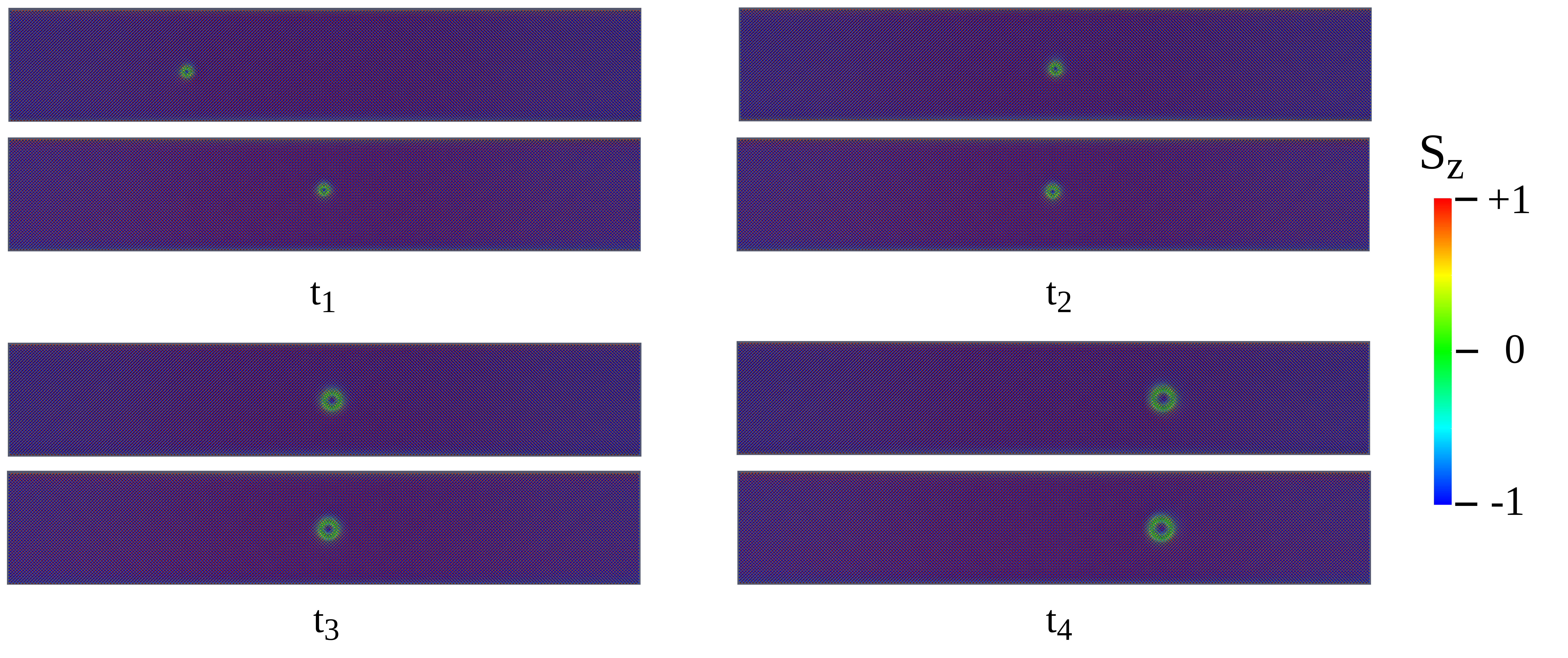}\label{F4a}}
	\subfigure[]{\includegraphics[width=3.65cm]{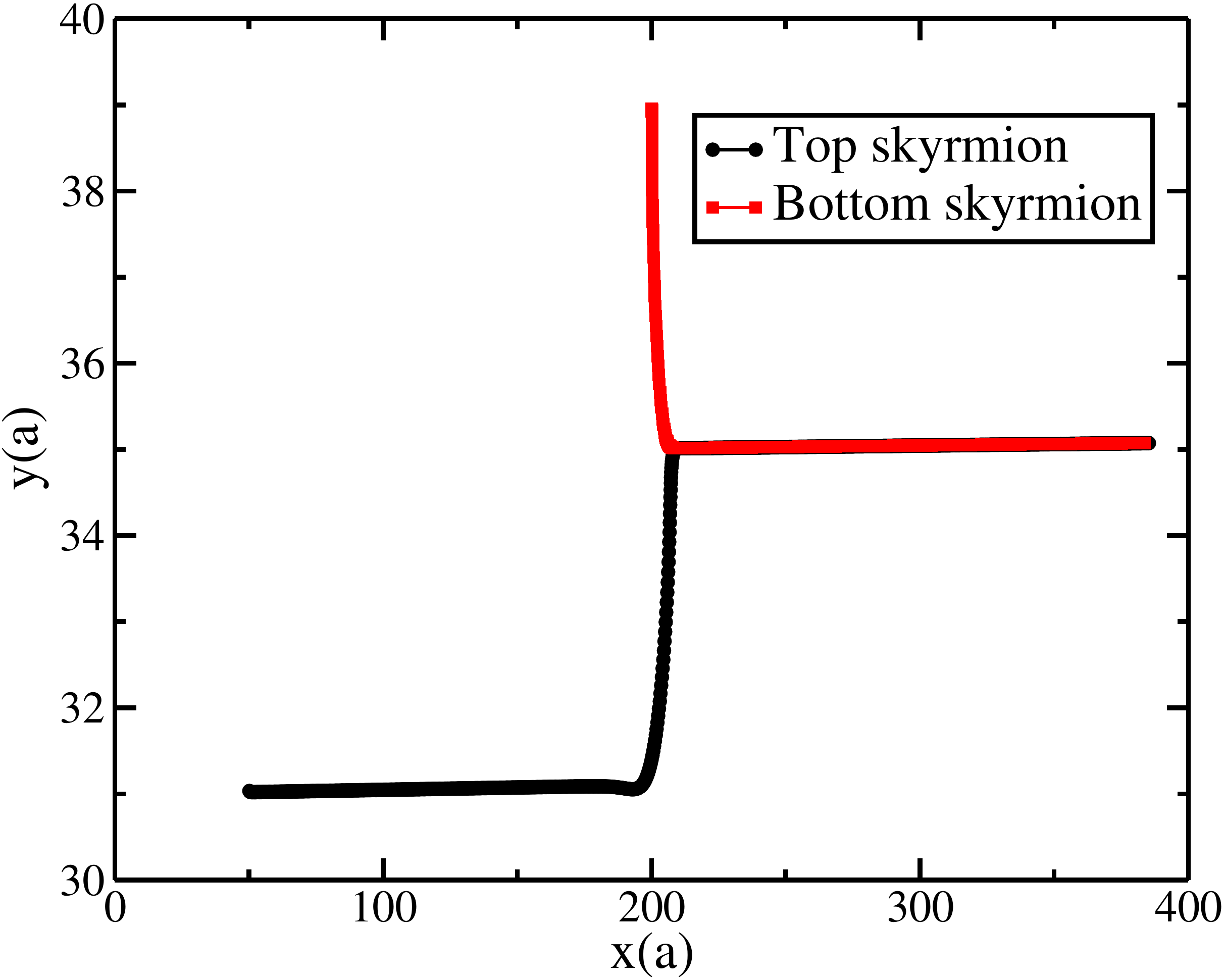}\label{F4b}}
	\subfigure[]{\includegraphics[width=3.5cm]{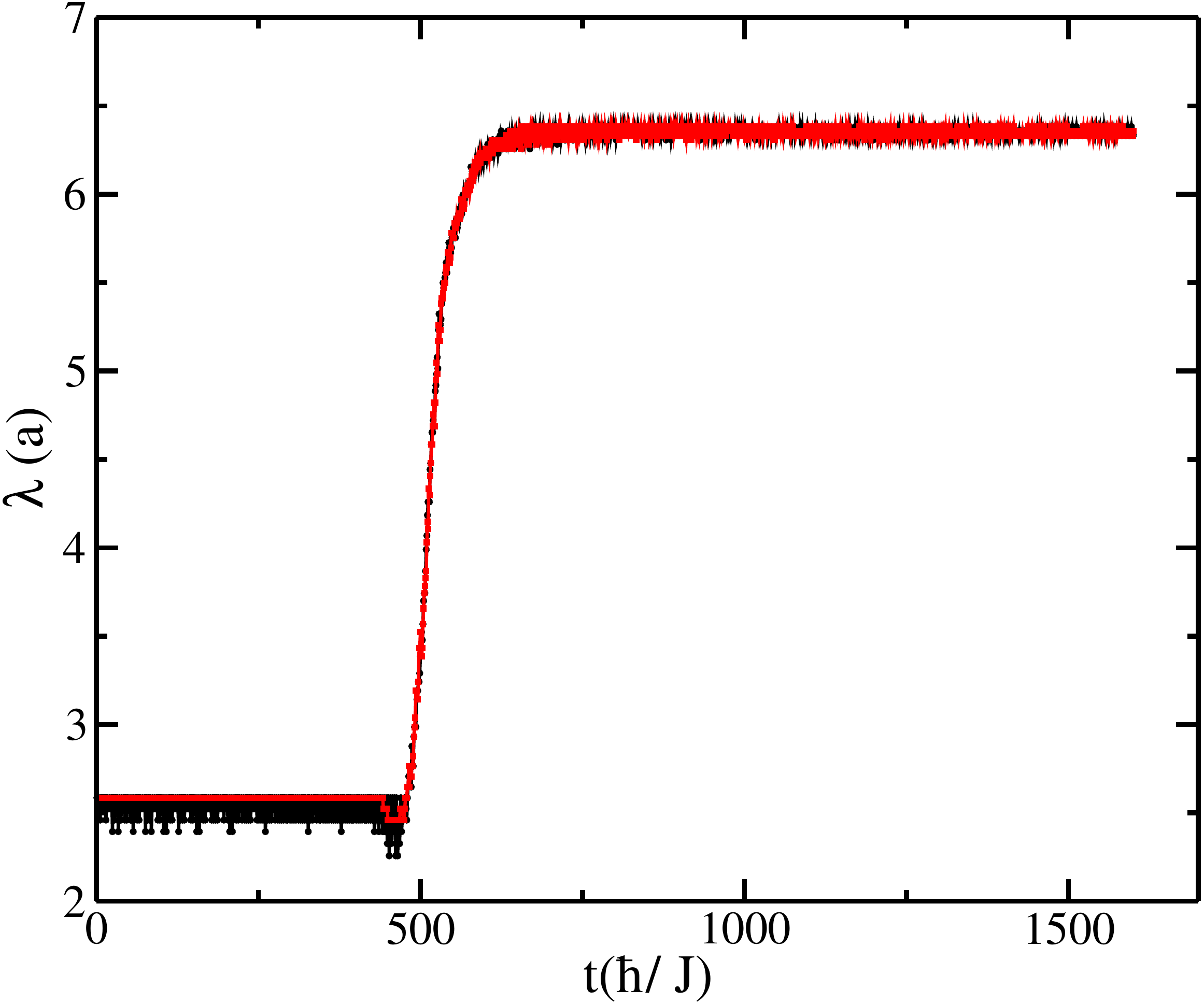}\label{F4c}}
	\vskip 0.1cm
	\subfigure[]{\includegraphics[width=6.30cm]{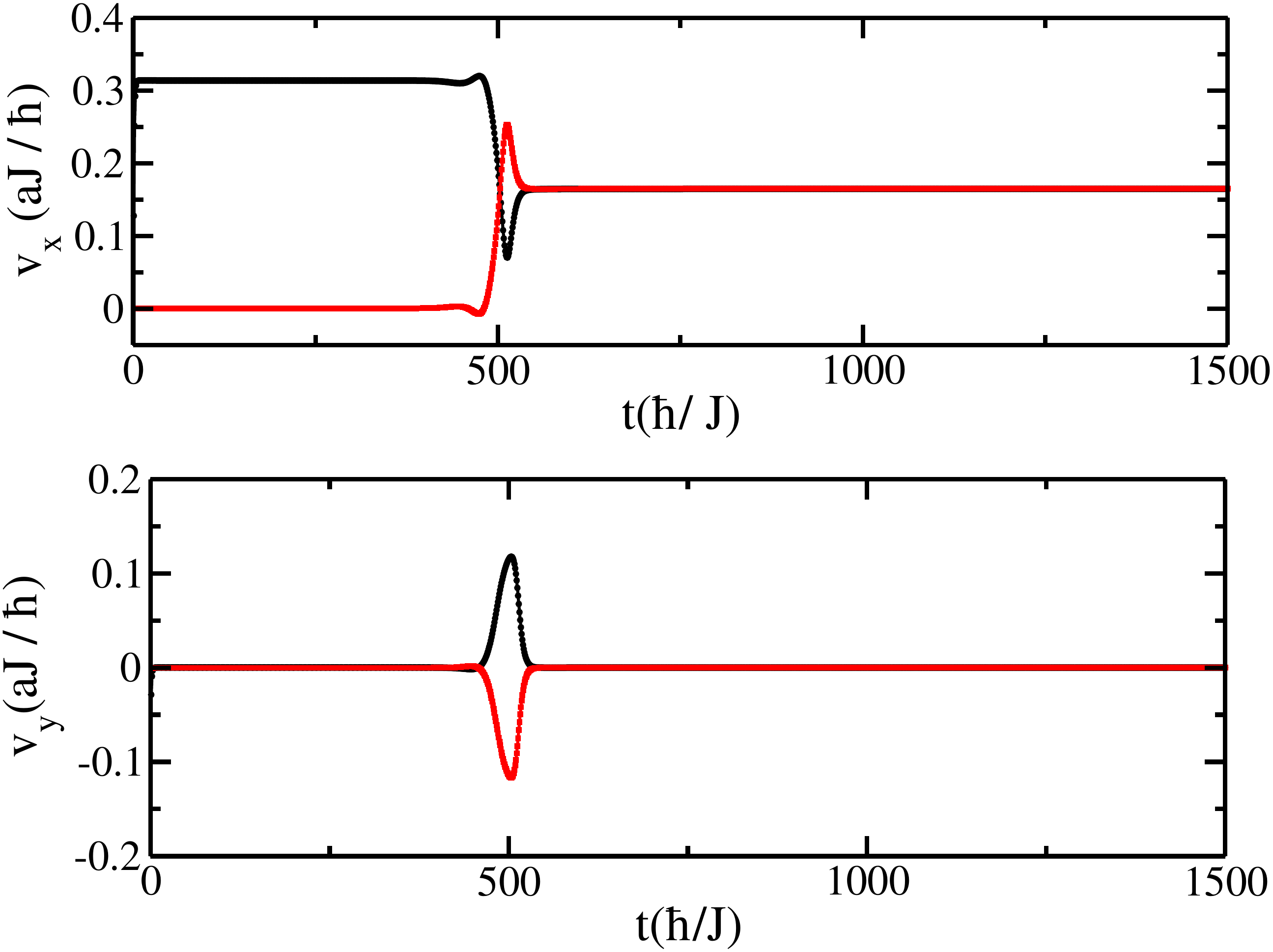}\label{F4d}}
	\caption{(Color online) (a) Snapshots showing time evolution of the skyrmion bound state (SBS) formation: at $t_{1}$ the top skyrmion moves to the right direction (along applied current), whereas at the bottom layer its counterpart is fixed at the middle of racetrack (no current is applied to this racetrack at all); at $t_{2}$ both skyrmions are in the same horizontal position, minimizing its spatial separation and yielding strong attraction between them, so that the formation of a SBS is achieved; at $t_{3}$ its is schematically shown how the SBS formation increases the size of both skyrmions; finally, at $t_{4}$ skyrmions forming the SBS displace together along the current direction. (b) The trajectory of skyrmions on their respective nanostripes. The black and red lines represent the paths of the upper/lower skyrmions. (c) Size of the skyrmions as a function of time. Their sizes increase considerably during SBS formation.  (d) The velocity components $v_{x}$ and $v_{y}$ of both skyrmions as a function of time.After SBS formation, both skyrmions displace together driven by the applied current. Namely, note how SBS formation is accompanied by strong and fast changes in the physical quantities above.}
	
	\label{figures4}
    \end{center}
\end{figure}

The possibility of nucleating a skyrmion bound-state (SBS) is very interesting because it allows moving one skyrmion by controlling the other one. Therefore, we have analyzed the influence of the external current ($j_e$) applied in the top layer and the initial vertical distance ($\Delta\,y$) in establishing the SBS. The main results are depicted in Fig. \ref{fig3New}, where pink and blue areas represent respectively the range of parameters to allow an SBS and the scattered skyrmion state (SSS). In this second case, the skyrmions move to each other, but the high current densities or big initial vertical distances do not allow the formation of a bound-state. Indeed, the maximum value for the vertical misalignment between the excitations must not exceed a few lattice spacing for the bound-state to occur. The critical current density for which the SSS appears decreases as $\Delta\,y$ increases (see the movies 1 and 2 available as supplementary material \cite{movies}). Therefore, it can be noticed that the increase of both current density and vertical distance diminishes the possibility of coupling the skyrmions. Regions with $j_e>0.11$ are not shown in Fig. \ref{fig3New} because in this case, the external current pulls the skyrmion in the top layer with a very large velocity in such a way that the SSS is obtained for any initial vertical distance.

We can now analyze the dynamic properties of the skyrmions when they form a bound-state. Figure \ref{F4a} shows some snapshots during the skyrmion motion. The up and down stripes in each snapshot represent the top and bottom layers, respectively. Indeed, the paths taken by each skyrmion during the entire process are shown in Fig. \ref{F4b}. The black and red lines represent, respectively, the trajectories of the upper and lower skyrmions. Note that while the horizontal distance between the skyrmions is large enough, only the top skyrmion moves. As they approach, the interaction between them becomes stronger and the bound-state is created. After that, the SBS persists to move in a rectilinear path. It can be observed that initially, the skyrmion radii are approximately $\lambda\approx2.5\,a$. However, when the SBS is nucleated, the skyrmion radii rapidly increase to $\lambda\approx 6\,a$ (See Fig. \ref{F4c}), due to the extra exchange energy coming from the interlayer coupling. Changes in the skyrmion radius were also observed for skyrmions in others systems such as ferromagnetic bilayer coupled by RKKY interaction \cite{Cacilhas, Deger}, artificial interlayer exchange interaction \cite{Vagson}, and antiferromagnetically coupled skyrmions in nanodots \cite{Fatouhi}. In addition to the radius, the skyrmion velocity is also affected when they are coupled. In this context, we have determined the velocity of the skyrmions in both layers as a function of time. Initially, because the electric current is applied just in the top layer, the horizontal component of the top skyrmion velocity before the coupling is constant $V_{x} \approx 0.31 aJ/\hbar$ (See the upper graph of the Fig. \ref{F4d} - black line), while the horizontal velocity of the skyrmion in the bottom layer is null (red line in Fig. \ref{F4d}). Nevertheless, our results suggest that after the creation of the bound-state of skyrmions, the new structure moves in a straight line with half the initial velocity value ($V_{x}^{SBS} \approx 0.16 aJ/\hbar$). Similar to what occurs for coupled ferromagnetic skyrmions, this behavior should be associated with an increase in the coupled skyrmion effective mass \cite{Vagson}. Finally, Fig. \ref{F4d} also evidences that the skyrmions acquire a non-zero velocity in the vertical direction ($y$-axis) during the formation of the bound-state. This vertical displacement happens due to their initial vertical misalignment, $\Delta y = 8a$. A noteworthy fact is that due to the particle-like nature of the skyrmions, the momentum is a conserved quantity during all stages, from the top excitation dynamics to the emergence and dynamics of the novel AFM bound-state of skyrmions.

\subsection{Bilayer with magnetic impurities}
 
The investigation of disordered systems plays an essential role in skyrmion dynamics because these imperfections can change drastically its movement. For example, magnetic defects can alter the SHE angle expressively in FM media \cite{Stier}. Additionally, there are some theoretical and experimental shreds of evidence that skyrmions can be created, annihilated, and pinned by magnetic impurities \cite{Woo1, Hrabec, Stier, Juge, Silva2019, DToscano}. In this way, after analyzing the dynamical properties of coupled skyrmions in a bilayer without impurities or defects, we study the dynamics of the SBS in a disordered system. In this context, we perform the simulations considering a system in which an SBS is an initial configuration. This initial configuration is achieved by considering a Belavin-Polyakov solution, Eq.(\ref{BPskyrmion}) \, in each racetrack with the skyrmions occupying the same horizontal position (each in its respective layer) and succeeding by the previously described relaxing process. The magnetic impurities are included in both layers by a random distribution of magnetic moments having modulus different from the spins in the racetracks without defects.

\begin{figure}[htb]
    \begin{center}
	\subfigure[]{\includegraphics[width=6.50cm]{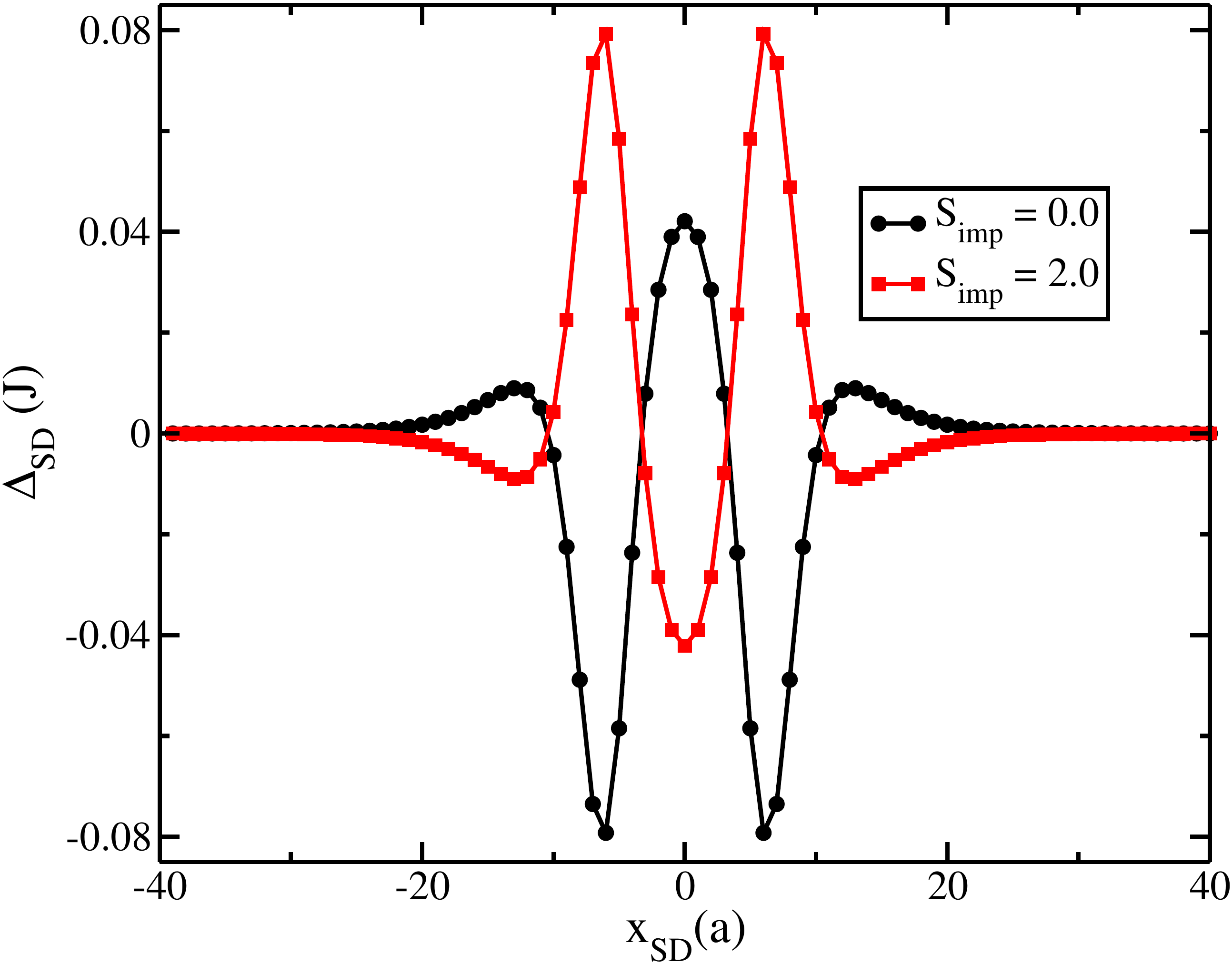}\label{F51}}
	\quad
	\subfigure[]{\includegraphics[width=6.50cm]{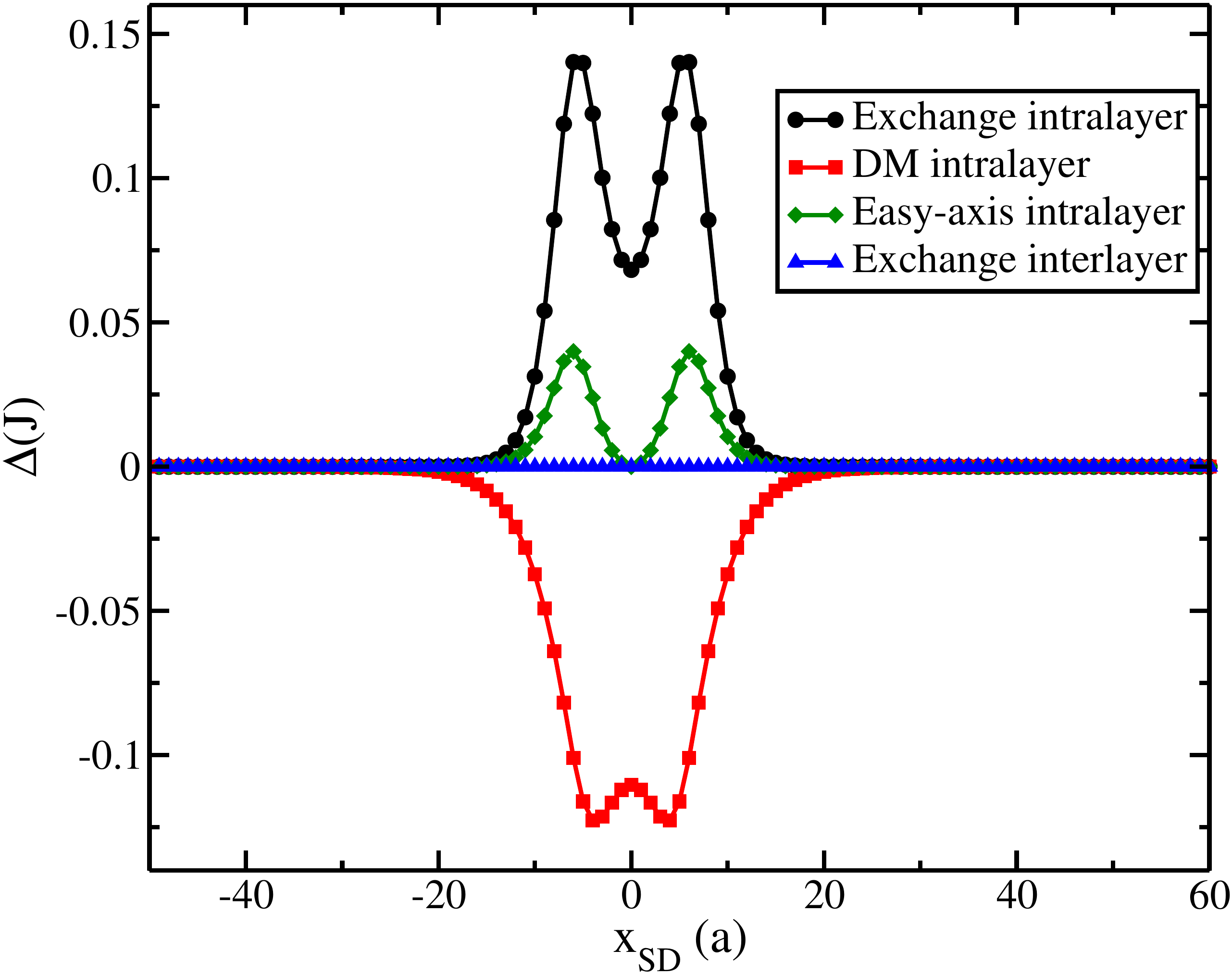}\label{F52}}
	\label{figures4}
	\caption{(Color online) (a) Interaction potential between the skyrmion bound state (SBS) and a lattice defect such as a magnetic impurity with $S_{imp} =2$ (red) and a nonmagnetic impurity, i.e., $S_{imp} =0$ (black). These types of impurities cause opposite effects on the SBS structure. (b) Individual contributions of the Hamiltonian terms for the SBS-magnetic impurity potential when $S_{imp} =2$: exchange intralayer, DM intralayer, easy-axis intralayer, and exchange interlayer are shown by black, red, green, and blue colors respectively. The sum of the contribution of all these individual interactions gives the red line in (a).}
    \end{center}
\end{figure}

To understand how the presence of a magnetic impurity affects the skyrmion dynamics, we have analyzed the interaction of the SBS with a single defect located at ($x_{imp}$, 35 a) of the upper nanostripe (there is no difference when considering such defect located at bottom nanotrack). Then, we calculate the energy of the whole system as a function of the impurity-SBS distance, $x_{SD}=x_{sk}-x_{imp}$.  Figure \ref{F51} presents the effective SBS-impurity interaction potential $\Delta_{SD} = E(x_{SD})-E(\infty)$, where $E(x_{SD})$ is the calculated total energy and $E(\infty)$ is the system energy when the distance between the SBS and the impurity is very large. The analysis of this figure reveals that, for $x_{SD}\gtrsim30\,a$, the skyrmions do not feel the presence of the magnetic impurity. However, for distances on the order of $x_{SD}\lesssim20\,a$, variations in the system's energy have been obtained, evidencing an impurity-SBS interaction. Moreover, the impurity-SBS interaction depends on the kind of defect present in the system. If the defect consists of a vacancy (the impurity is not a magnetic atom), represented by $S_{imp}=0$ (black dots in Fig. \ref{F51}), the skyrmions are initially repelled by the defect. When $x_{SD}\lesssim10\,a$, the skyrmions experience an attraction until their cores are at a distance $x_{SD}\lesssim8\,a$ from the magnetic impurity, when the system energy reaches the minimum energy configuration. Because the maximum energy occurs when the skyrmion core is just at the defect position, the vacancy repels the SBS. Analogous and opposite behavior is obtained when the impurity consists of an extra magnetic atom with a higher magnetic moment, here represented for $S_{imp}=2$ (red squares in Fig. \ref{F51}). In this case, $\Delta_{SD}$ reaches a minimum when the skyrmions are at the impurity position. Above described results evidence that vacancies and extra magnetic atoms can act as skyrmion scattering and pinning centers. Although this work reports only the cases with $S_{imp}=0$ and $S_{imp}=2$, the magnitude of the impurities in the interval $0.0 \leq S_{imp} \leq 2.0$ were also investigated. Nonetheless, the same qualitative behavior is observed for other $S_{imp}$ values, and only quantitative changes are observed for the amplitude of $\Delta_{SD}$.

\begin{figure}[htb]
	\begin{center}
		\includegraphics[width=8.5cm]{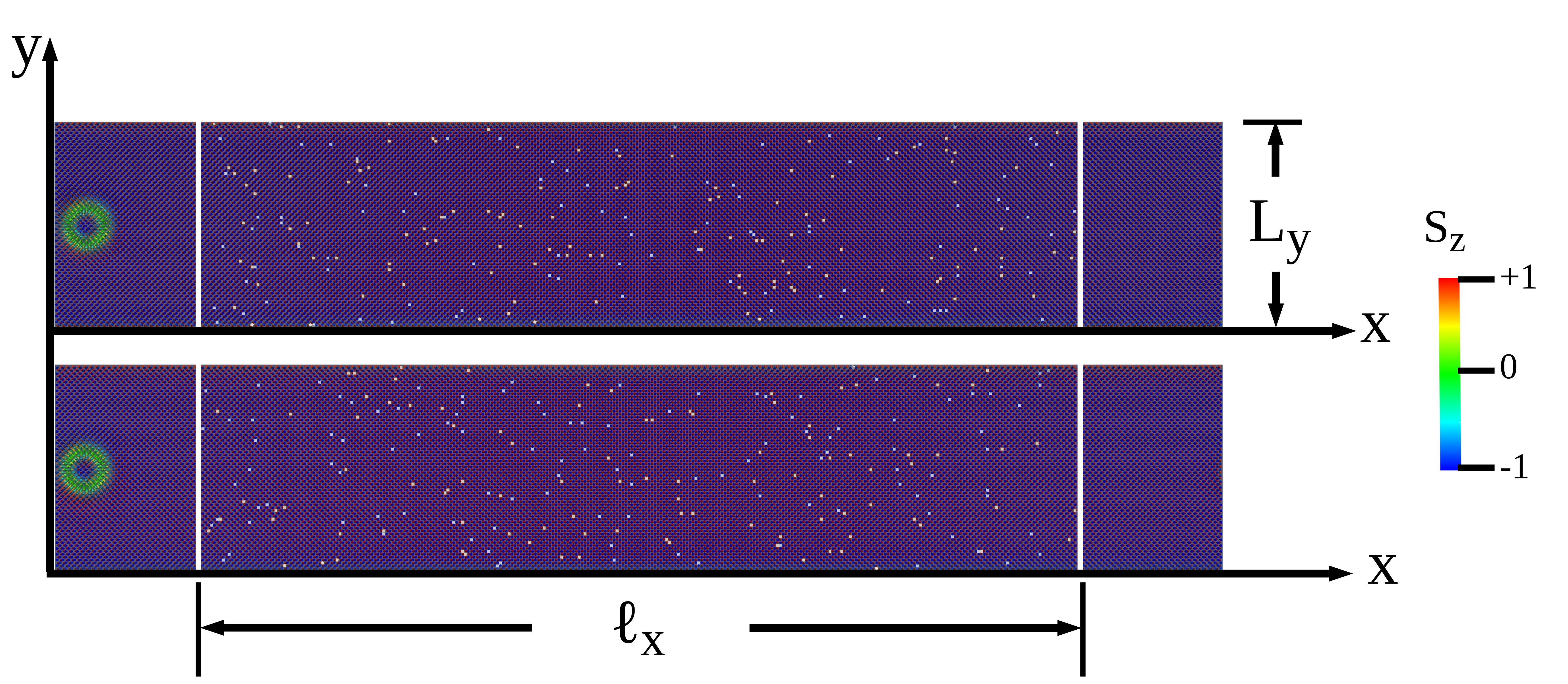}
		\caption{(Color online) Side-by-side view of the racetrack with randomly distributed impurities over regions of length $\ell_x=300a$. Skyrmion patterns shown form a SBS and they will be driven towards such inhomogeneous regions.}
		\label{dirty}
	\end{center}
\end{figure}

It is worth noticing that the inclusion of magnetic defects is interpreted as a simultaneous disorder in all magnetic couplings, which are scaled by the defective magnetic moment modulus. Thus, the potential $\Delta_{SD}$ consists of the sum of the intralayer exchange, anisotropy, DM, and interlayer exchange interactions. Therefore, it is possible to determine the impurity-induced changes in the contribution of each magnetic interaction considered in Eq. \eqref{hamiltonian}. The obtained results for $S_{imp} = 2$ are presented in Fig. \ref{F52}, where black circles, red squares, green diamonds, and blue triangles depict the behavior of the intralayer exchange, DM, easy-axis anisotropy, and the interlayer exchange interactions respectively. It can be noticed that the inclusion of an extra magnetic atom with a higher magnetic moment yields a local variation in the intralayer exchange and easy-axis anisotropy constants that present a repulsive character. On the other hand, this kind of magnetic impurity leads to an increase in the DM coupling, evidencing that the resulting attractive potential observed in Fig. \ref{F51} (for the case where $S_{imp} =2$) is originated from the impurity-induced changes in the DM interaction. These results are in good agreement with those reported in the Ref. \cite {DToscano}, where the authors analyzed local variations in materials parameters of a single AFM nanostripe, observing both attractive or repulsive potentials due to variations in the local magnetic coupling interactions.

After describing the SBS-impurity interaction, we can study how the SBS dynamic is modified when passing through a region with a certain concentration $\rho$ of magnetic impurities. These magnetic defects are randomly distributed in each layer, inside a region containing $\ell_{x}=300 a$ spins in the horizontal direction and $L_{y}=70a$ in the vertical direction (See Fig. \ref{dirty}). Thus, in each layer, we compute a number of defective spins as $N_{d} = \rho \ell_{x}Ly$. Initially, the SBS is placed at $(20 a, 35 a)$, when a spin-polarized current density $j_{e}=0.10$ starts to move the SBS to the region with magnetic defects. This procedure was performed in 20 independent simulations considering different distributions of magnetic impurities along the nanotracks. All performed simulations yield similar results, and from now on, we present the obtained dynamical parameter for just one of the considered samples.
\begin{figure}[htb]
    \begin{center}
	\subfigure[]{\includegraphics[width=6.0cm]{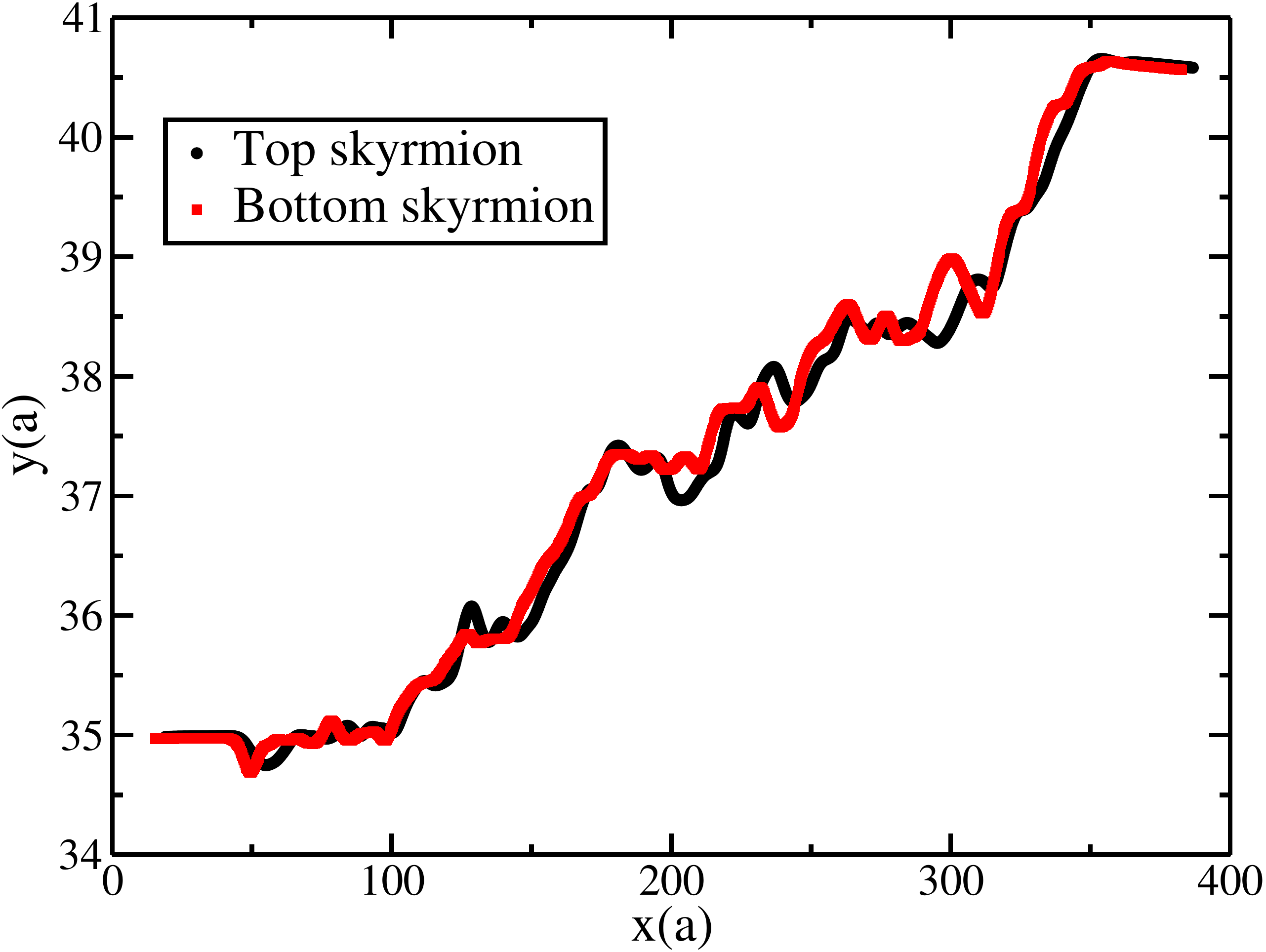}\label{F72}}
	\subfigure[]{\includegraphics[width=6.00cm]{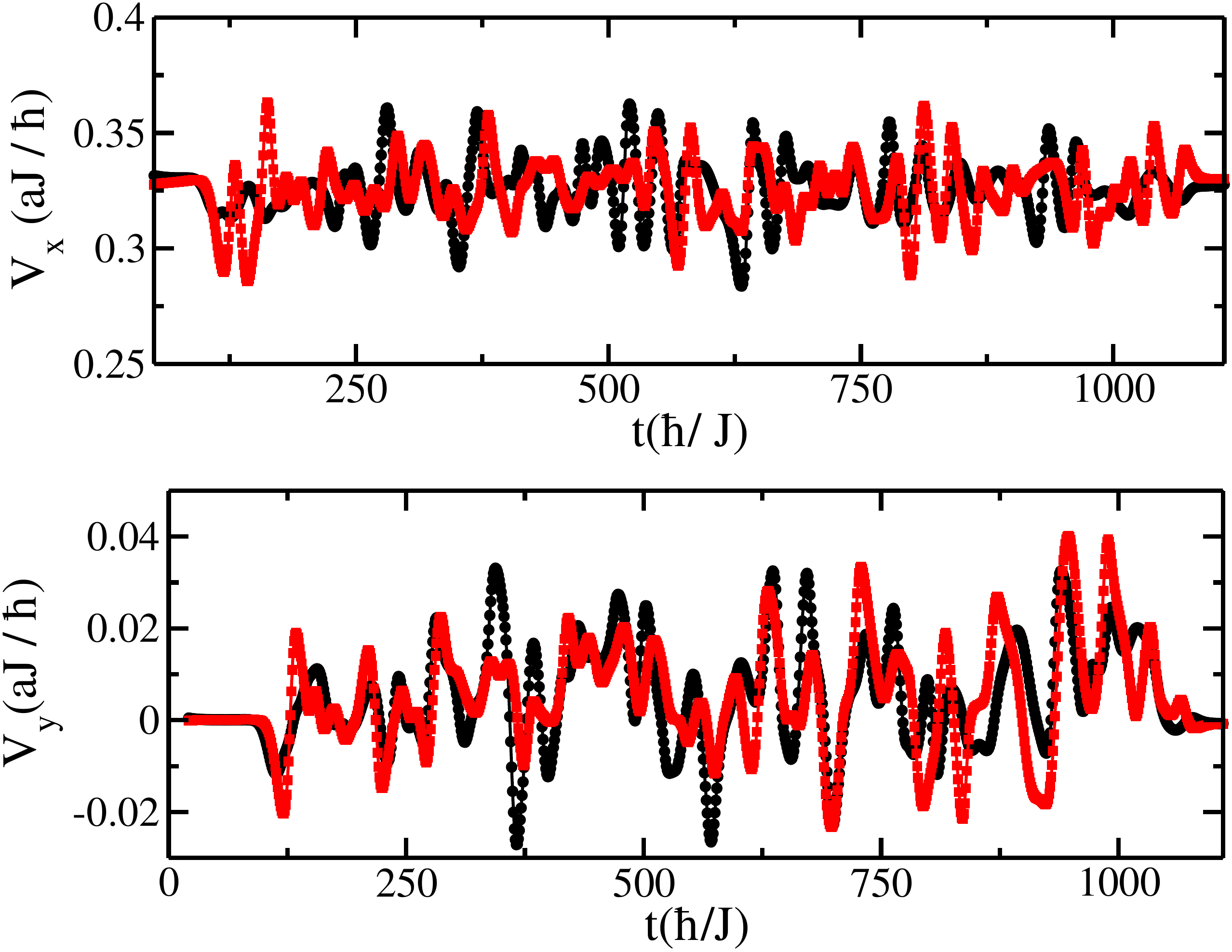}\label{F73}}
	
	\label{figures7}
	\caption{(Color online) How skyrmion dynamics is affected by impurities. (a) Typical trajectory of a SBS with magnetic impurity ($S_{imp}=2$) concentration of $\rho=1\%$. Due to such impurities, skyrmion trajectory becomes very complex and it is slightly deflected from the rectilinear motion. (b) Horizontal and vertical skyrmions velocities are affected by impurities along their motion. Compare with the case of a clean bilayer, Fig. \ref{F4d}.}
    \end{center}
\end{figure}

Figure \ref{F72} presents the typical trajectory of the SBS in a bilayer with concentration $\rho=1\%$ of extra magnetic atoms (impurities with $S_{imp} = 2.0$), placed in the region $50a \leq x \leq 350 a$. Black and red-dashed lines respectively depict the position as a function of the time for a skyrmion in the top and bottom layers. It can be noticed that in the region $x < 50 a$ and $x>350a$, the SBS moves in a straight line. Nevertheless, due to the SBS-magnetic impurity interaction, its trajectory randomly deviates from the rectilinear movement when it enters the region with magnetic impurities. After crossing the region with defects, the skyrmion trajectory is displaced in a few lattice spacing along the vertical direction. A similar result was reported in Ref. \cite{Silva2021}, where the author analyzed the dynamics of an AFM bimeron in a track with magnetic impurities.  The effects of the magnetic impurities in the skyrmion dynamics can also be evidenced from the analysis of the SBS velocity as a function of time (See Fig. \ref{F73}). One can see that when the SBS moves in the region free of impurities, its velocity has only a horizontal component, $V_{x} \approx 0.330 a\hbar/J$. The SBS-impurity interaction becomes evident when the coupled skyrmions enter the region with defects, where the pinning (or scattering) of SBS by the magnetic defects yields fluctuations in the SBS  velocity. The components $V_{x}$ and $V_{y}$ of the top and bottom skyrmions are depicted in Fig. \ref{F73}, where black and red-dashed lines, respectively. Finally, we have also performed micromagnetic simulations for different impurity concentrations. The obtained results reveal that for $\rho>1\%$ there is the appearance of clusters of magnetic impurities, which can pin the skyrmion in the bottom layer, and the SBS is eventually annihilated (Movies 3 and 4 in the supplementary material presents its complete dynamics in more detail \cite{movies}).

 \section{Conclusions}

In summary, we have investigated an antiferromagnetic bilayer racetrack with a small ferromagnetic coupling between the layers, which are separated by a height offset. Each layer comprises a single skyrmion pattern. We have shown that such skyrmions experience a strong attraction whenever they are relatively close. In this case, they behave as they formed a bound state. For instance, if one applies current along the top layer, both skyrmions displace together driven by the spin transfer torque experienced by the top skyrmion. In the idealized case of a clean racetrack, skyrmion bound state displaces in rectilinear motion. Whenever magnetic inhomogeneity is taken into consideration, we have seen that although skyrmion dynamics is affected (e.g. its trajectory is deflected and present local sharp variations due to strong scattering at impurity sites) the bound state is sufficiently robust to pass through a region of randomly distributed impurities. We hope our findings may encourage further studies concerning skyrmion bound state as useful candidates for information carriers in topological spintronic research. 

 \section{Acknowledgments}
 The authors thank the Brazilian agencies CAPES (Finance Code 001), CNPq, FAPEMIG and FAPES for financial support.


\begin{thebibliography}{99}

\bibitem{Skyrme1} T. H. R. Skyrme, Proc. R. Soc. A \textbf{260}, 127 (1961).
\bibitem{Skyrme2} T. H. R. Skyrme, Nucl. Phys. \textbf{31}, 556 (1962).
\bibitem{BP}  A. A. Belavin and A.M. Polyakov, JETP Lett. \textbf{22}, 245 (1975).
\bibitem{Dzyaloshinskii} I. Dzyaloshinskii, J. Phys. Chem. Solids \textbf{4}, 241 (1958).
\bibitem{Moriya} T. Moriya, Phys. Rev. \textbf{120}, 91 (1960).
\bibitem{ANBogdonov} A. N. Bogdanov, D. A. Yablonski, Sov. Phys. JETP \textbf{68}, 101 (1989).
\bibitem{ABogdanov} A. Bogdanov, A. Hubert, J. Magn. Magn. Mat. \textbf{138}, 255 (1994).
\bibitem{Robler} U. K. R\"{o}{\ss}ler, A. N. Bogdanov, C. Pfleiderer, Nat. \textbf{442}, 797 (2006).
\bibitem{Heinze} S. Heinze, K. von Bergmann, M. Menzel, J. Brede, A. Kubetzka, R. Wiesendanger, G. Bihlmayer, S. Bl\"{u}gel, Nat. Phys. \textbf{7}, 713 (2011).
\bibitem{Wiesendanger} R. Wiesendanger, Rev. Mater. \textbf{1}, 16044 (2016).
\bibitem{Soumyanarayanan} A. Soumyanarayanan, M. Raju, A. L. Gonzalez Oyarce, A. K. C. Tan, M. -Y. Im, A. P. Petrovi\'{c}, P. Ho, K. H. Khoo, M. Tran, C. K. Gan, F. Ernult, C. Panagopoulos, Nat. Mater. \textbf{16}, 898 (2017).
\bibitem{Iwasaki} J. Iwasaki, M. Mochizuki, N. Nagaosa, Nat. Commun. \textbf{4}, 1463 (2013).
\bibitem{Romming} N. Romming, C. Hanneken, M. Menzel, J. Bickel, B. Wolter, K. von Bergmann, A. Kubetzka, R. Wiesendanger, Sci. \textbf{341}, 636 (2013).
\bibitem{Chouk} A. Derras-Chouk, E. M. Chudnovsky, D. A. Garanin, Phys. Rev. B \textbf{98}, 024423 (2018).
\bibitem{Heil} B. Heil, A. Rosch, J. Masell, Phys. Rev. B \textbf{100}, 134424 (2019).
\bibitem{XZhang3} X. Zhang, Y. Zhou, K. M. Song, T. -E. Park, J. Xia, M. Ezawa, X. Liu, W. Zhao, G. Zhao, S. Woo, J. Phys.: Condens. Matter \textbf{32}, 143001 (2020).
\bibitem{Brandao} J. Brand\~{a}o, D. A. Dugato, R. L. Seeger, J. C. Denardin, T. J. A. Mori, J. C. Cezar, Sci. Rep. \textbf{9}, 4144 (2019).
\bibitem{Woo1} S. Woo, K. Litzius, B. Kr\"{u}ger, M. -Y. Im, L. Caretta, K. Richter, M. Mann, A. Krone, R. M. Reeve, M. Weigand, P. Agrawal, I. Lemesh, M. -A. Mawass, P. Fischer, M. Kl\"{a}ui, G. S. D. Beach, Nat. Mater. \textbf{15}, 501 (2016).
\bibitem{Yu} X. Z. Yu, N. Kanazawa, W. Z. Zhang, T. Nagai, T. Hara, K. Kimoto, Y. Matsui, Y. Onose, Y. Tokura, Nat. Commun. \textbf{3}, 988 (2012).
\bibitem{JZang} J. Zang, M. Mostovoy, J. H. Han, N. Nagaosa, Phys. Rev. Lett. \textbf{107}, 136804 (2011).
\bibitem{Jiang} W. Jiang, X. Zhang, G. Yu, W. Zhang, M. B. Jungfleisch, J. E. Pearson, O. Heinonen, K. L. Wang, Y. Zhou, A. Hoffmann, S. G. E. te Velthuis, Nat. Phys. \textbf{13}, 162 (2017).
\bibitem{Litzius} K. Litzius, I. Lemesh, B. Kr\"{u}ger, P. Bassirian, L. Caretta, K. Richte, F. B\"{u}ttner, K. Sato, O. A. Tretiakov, J. F\"{o}rster, R. M. Reeve, M. Weigand, I. Bykova, H. Stoll, G. Sch\"{u}tz, G. S. D. Beach, M. Kl\"{a}ui, Nat. Phys. \textbf{13}, 170 (2017).
\bibitem{Toscano} D. Toscano, J. P. A. Mendona, A. L. S. Miranda, C. I. L. de Araujo, F. Sato, P. Z. Coura, S. A. Leonel, J. Magn. Magn. Mater. \textbf{504}, 166655 (2020).
\bibitem{Silva} R. C. Silva, R. L. Silva, A. R. Pereira,  J. Phys.: Condens. Matter \textbf{33}, 105802 (2021).
\bibitem{Kolesnikov} A. G. Kolesnikov, M. E. Stebliy, A. S. Samardak, A. V. Ognev, Sci. Rep. \textbf{8}, 16966 (2018).
\bibitem{XZhang1} X. Zhang, Y. Zhou, M. Ezawa, Nat. Commun. \textbf{7}, 10293 (2016).
\bibitem{Koshibae} W. Koshibae, N. Nagaosa, Sci. Rep. \textbf{7}, 42645 (2017).
\bibitem{Hrabec} A. Hrabec, J. Sampaio, M. Belmeguenai, I. Gross, R. Weil, S. M. Ch\'{e}rif, A. Stashkevich, V. Jacques, A. Thiaville, S. Rohart, Nat. Commun. \textbf{8}, 15765 (2017).

\bibitem{Vagson} S. Vojkovic, R. Cacilhas, A. R. Pereira, D. Altbir , A. S. N\'{u}\~{n}ez and V. L. Carvalho-Santos, Nanotechnology \textbf{32}, 175702 (2021).

\bibitem{Cacilhas}
R. Cacilhas, V. L. Carvalho-Santos, S. Vojkovic, E. B. Carvalho, A. R. Pereira, \'A. S. N\'u\~nez, and D. Altbir, Appl. Phys. Lett.
\textbf{113}, 212406 (2018)

\bibitem{Smejkal} L. {\v S}mejkal, Y. Mokrousov, A. H. MacDonald, Nat. Phys. \textbf{14}, 242 (2018).

\bibitem{XZang-JAP}
X. Liang, J. Xia, X. Zhang, M. Ezawa, O. A.Tretiakov, X. Liu, L. Qiu, G. Zhao, and Y. Zhou, Appl. Phys. Lett. \textbf{119}, 062403 (2021). 

\bibitem{Barker} J. Barker, O. A. Tretiakov, Phys. Rev. Lett. \textbf{116}, 147203 (2016).
\bibitem{XZhang2} X. Zhang, Y. Zhou, M. Ezawa, Sci. Rep. \textbf{6}, 24795 (2016).
\bibitem{Jin} C. Jin, C. Song, J. Wang, Q. Liu, Appl. Phys. Lett. \textbf{5}, 9400 (2016).
\bibitem{Woo2} S. Woo, K. M. Song, X. Zhang, Y. Zhou, M. Ezawa, X. Liu, S. Finizio, J. Raabe, N. J. Lee, S. -I. Kim, S. -Y. Park, Y. Kim, J. -Y. Kim, D. Lee, O. Lee, J. W. Choi, B. -C. Min, H. C. Koo, J. Chang, Nat. Commun. \textbf{9}, 959 (2018).
\bibitem{Chen} K. Chen, D. Lott, A. Philippi-Kobs, M. Weigand, C. Luo, F. Radu, Nanoscale \textbf{12}, 18137 (2020).
\bibitem{Gao} S. Gao, H. D. Rosales, F. A. G. Albarrac\'{i}n, V. Tsurkan, G. Kaur, T. Fennell, P. Steffens, M. Boehm, P. {\v C}erm\'{a}k, A. Schneidewind, E. Ressouche, D. C. Cabra, C. R\"{u}egg, O. Zaharko, Nat. \textbf{586}, 37 (2020).
\bibitem{Raicevic} I. Raicevic, D. Popovic, C. Panagopoulos, L. Benfatto, M. B. Silva Neto, E. S. Choi, T. Sasagawa, Phys. Rev. Lett. \textbf{106}, 227206 (2011).
\bibitem{Burgler} D. E. B\"{u}rgler, P. Gr\"{u}nberg, S. O. Demokritov, M. T. Johnson, Handbook of Magnetic Materials, ed. K. H. J. Buschow, 13 (Amsterdam: Elsevier) (2001).
\bibitem{Orozco} A. Orozco, S. B. Ogale, Y. H. Li, P. Fournier, Eric Li, H. Asano, V. Smolyaninova, R. L. Greene, R. P. Sharma, R. Ramesh, T. Venkatesan, Phys. Rev. Let. \textbf{83}, 1680 (1999).
\bibitem{Bruno} P. Bruno, Phys. Rev. B \textbf{52}, 411 (1995).
\bibitem{SHYang} S. -H. Yang, K. -S. Ryu, S. Parkin, Nat. Nanotech. \textbf{10}, 221 (2015).
\bibitem{GChen} G. Chen, T. Ma, A. T. N'Diaye, H. Kwon, C. Won, Y. Wu, A. K. Schmid, Nat. Commun. \textbf{4}, 2671 (2013).
\bibitem{Dupe} B. Dup{\'{e}}, G. Bihlmayer, M. B{\"{o}}ttcher, S. Bl{\"{u}}gel, S. Heinze, Nat. Commun. \textbf{7}, 11779 (2016).
\bibitem{Luchaire} C. Moreau-Luchaire, C. Moutafis, N. Reyren, J. Sampaio, C. A. F. Vaz, N. Van Horne, K. Bouzehouane, K. Garcia, C. Deranlot, P. Warnicke, P. Wohlh{\"{u}}ter, J. -M. George, M. Weigand, J. Raabe, V. Cros, A. Fert, Nat. Nanotech. \textbf{11}, 444 (2016). 
\bibitem{Matsuno} J. Matsuno, N. Ogawa, K. Yasuda, F. Kagawa, W. Koshibae, N. Nagaosa, Y. Tokura, M. Kawasaki, Sci. Adv. \textbf{2}, e1600304 (2016).
\bibitem{Parkin} S. S. P. Parkin, N. More, K. P. Roche, Phys. Rev. Lett. \textbf{64}, 2304 (1990).
\bibitem{Parkin2} S. S. P. Parkin, C. Kaiser, A. Panchula, P. M. Rice, B. Hughes, M. Samant, S. -H. Yang, Nat. Mat. \textbf{3}, 862 (2004).
\bibitem{Yuasa} S. Yuasa, T. Nagahama, A. Fukushima, Y. Suzuki, K. Ando, Nat. Mat. \textbf{3}, 868 (2004).
\bibitem{Nunn} Z. R. Nunn, C. Abert, D. Suess, E. Girt, Sci. Adv. \textbf{6}, eabd8861 (2020).
\bibitem{Rohart} S. Rohart, A. Thiaville, Phys. Rev. B \textbf{88}, 184422 (2013).
\bibitem{Yang} H. Yang, A. Thiaville, S. Rohart, A. Fert, M. Chshiev, Phys. Rev. Lett. \textbf{115}, 267210 (2015).
\bibitem{Ubiergo} R. Jaeschke-Ubiergo, A. S. Nunez, Annals of Phys. \textbf{405}, 29 (2019).
\bibitem{Landau} L. D. Landau, E. M. Lifshitz, Phys. Zs. Sowjet. \textbf{8}, 153 (1935).
\bibitem{Gilbert} T. L. Gilbert, IEEE Trans. Magn. \textbf{40}, 3443 (2004).
\bibitem{Zhang-Li} S. Zhang, Z. Li, Phys. Rev. Lett. \textbf{93}, 127204 (2004).
\bibitem{Moutafis} C. Moutafis, S. Komineas, J. A. C. Bland, Phys. Rev. B \textbf{79}, 224429 (2009).
\bibitem{Shen} L. Shen, J. Xia, X. Zhang, M. Ezawa, O. A. Tretiakov, X. Liu, G. Zhao, Y. Zhou, Phys. Rev. Lett. \textbf{124}, 037202 (2020).
\bibitem{movies} Watch the video files in the online version of this paper. The geometric parameters of AFM bilayer  is L$_{x}$ = 400 \textit{a} and L$_{y}$ = 70 \textit{a}. In all cases, a spin-polarized current is applied in the horizontal direction of the upper layer. mv1.mp4 and mv2.mp4 show, respectively, the formation of the skyrmion bound-state (SBS) and a skyrmion scattering (SSS) processes during the movement of the skyrmion located at the top racetrack. mv3.mp4 and mv4.mp4 present the SBS dynamics through a region with magnetic defects with different impurities concentrations $\rho$. In mv3.mp4, we have used $\rho=1\%$ while in mv4.mp4, $\rho = 2\%$ with randomly magnetic impurities with $S_{imp}=2$.
\bibitem{Stier} M. Stier, R. Strobel, S. Krause, W. H{\"a}usler, M. Thorwart, Phys. Rev. B \textbf{103}, 054420 (2021).
\bibitem{Deger} C. Deger, I. Yavuz, and F. Yildiz, J. Magn. Magn. Mater. \textbf{489}, 165399 (2019).
\bibitem{Fatouhi} M. Fattouhi, M. Y. E. Hafidi, and M. E. Hafidi, Phys. Lett. A \textbf{384}, 126260 (2020).
\bibitem{Juge} R. Juge, S. G. Je, D. S. Chaves, L. D. Buda-Prejbeanu, J. Pena-Garcia, J. Nath, I. M. Miron, K. G. Rana, L. Aballe, M. Foerster, F. Genuzio, T. O. Mentes, A. Locatelli, F. Maccherozzi, S. S. Dhesi, M. Belmeguenai, Y. Roussigne, S. Auffret, S. Pizzini, G. Gaudin, J. Vogel, O. Boulle, Phys. Rev. Applied \textbf{12}, 044007 (2019).
\bibitem{Silva2019} R. L. Silva, R. C. Silva, A. R. Pereira, and W. A. Moura-Melo, J. Phys.: Condens. Matter \textbf{31}, 225802 (2019).
\bibitem{DToscano} D. Toscano, I. A. Santece, R. C. O. Guedes, H. S. Assis, A. L. S. Miranda,  C. I. L. de Araujo, F. Sato,  P. Z. Coura, S. A. Leonel, J. Appl. Phys. \textbf{127}, 193902 (2020).
\bibitem{Silva2021} R. L. Silva, Phys. Lett. A \textbf{403}, 127399 (2021).

\end{thebibliography}
\end{document}